# Laser patterned polymer/nanotube composite electrodes for nanowire transistors on flexible substrates

Kiron Prabha Rajeev; Michael Beliatis; Stamatis Georgakopoulos; Vlad Stolojan; John Underwood and Maxim Shkunov

Advanced Technology Institute, Electrical and Electronic Engineering, University of Surrey, Guildford GU2 7XH, United Kingdom

## Abstract

Fabrication techniques such as laser patterning offer excellent potential for low cost and large area device fabrication. Conductive polymers can be used to replace expensive metallic inks such as silver and gold nanoparticles for printing technology. Electrical conductivity of the polymers can be improved by blending with carbon nanotubes. In this work, formulations of acid functionalised multiwall carbon nanotubes (f-MWCNT) and poly (ethylenedioxythiophene) [PEDOT]: polystyrene sulphonate [PSS] were processed, and thin films were prepared on plastic substrates. Conductivity of PEDOT: PSS increased almost four orders of magnitude after adding f-MWCNT. Work function of PEDOT: PSS/f-MWCNT films was ~ 0.5eV higher as compared to the work function of pure PEDOT: PSS films, determined by Kelvin probe method. Field-effect transistors source-drain electrodes were prepared on PET plastic substrates where PEDOT: PSS/f-MWCNT were patterned using laser ablation at 44mJ/pulse energy to define 36μm electrode separation. Silicon nanowires were deposited using dielectrophoresis alignment technique to bridge the PEDOT: PSS/f-MWCNT laser patterned electrodes. Finally, top-gated nanowire field effect transistors (FET) were completed by depositing parylene C as polymer gate dielectric and gold as the top-gate electrode. Transistor characteristics showed p-type conduction with excellent gate electrode coupling, with an ON/OFF ratio of ~ 200. Thereby, we demonstrate the feasibility of using high workfunction, printable PEDOT: PSS/MWCNT composite inks for patterning source/drain electrodes for nanowire transistors on flexible substrates.

## Introduction

Solution processed nanoelectronic materials in the recent years have opened up possibilities for printed electronics applications such as sensors[1], flexible displays[2], energy harvesting piezoelectric devices[3]. High performance transistors are required for displays and sensors, whereas dense nanomaterials coverage is request for energy harvesting, and, in all cases, low cost deposition and structuring techniques are essential for the devices fabrication. Conventional photo-lithographic fabrication provides high resolution electrode deposition, but it does not fit well with high throughput processes[4]. Low cost and efficient technologies should be favoured in the patterning of electrodes such as ink-jet printing[5], laser patterning[6], soft lithography[7]. In particular, laser-based techniques to pattern the electrodes can be cost effective and efficient. The laser ablation technique is a one step process which can be applied to metals as well as polymers, without the necessity for any treatments such as etching and striping[8]. Efficient material removal for electrode patterning can be achieved by high intensity pulsed lasers. The induced laser pulse breaks the chemical bonds and removes the unwanted material from the substrate, for pulse with delivered energy above the binding energy of the molecules within the films on the substrate. The material can be removed with one or multiple laser pulses depending on the absorption properties of the coating and base substrate material, chemical structure of the films, film layer thickness and laser irradiation power density[8].

The use of solution processed highly conductive materials such as metal nanoparticle inks or polymers as electrodes for field effect transistors (FET), can significantly reduce the overall cost of fabrication. The deposition of such conductive materials can be achieved by large area printing techniques, which do not require special vacuum-based deposition equipment such as evaporation or sputtering. In the recent years, various solution processed conductive inks for printing technology have been demonstrated. The choice of metallic inks for printing technologies is relatively limited, and only gold, silver and copper inks are currently available. Gold inks offer high workfunction, close to 5eV, however, they are prohibitively expensive. Copper inks have lower workfunction, and printed layers are typically not very stable in ambient conditions. Air stable copper inks with antioxidants have been demonstrated recently, but higher temperature annealing is required for sintering[9]. Among only few available metallic nanoparticle inks, silver nanoparticle ones are the most common, however, the workfunction of printed layers (~ 4.5eV) is not high enough to provide high quality ohmic contact to many hole transport organic semiconductors, and solution processable silicon nanowires. The need for higher workfunction inks has led to the development of hybrid inks such as silver/organic electronic acceptors with reported workfunction value of ~5.1eV [10]. Whereas, conducting polymers, including PEDOT: PSS, poly (3, 4-ethelenedioxythiophene): poly (styrenesulphonate), offer higher workfunction $\phi$ of ~5.1eV [11], and PEDOT:PSS has been widely used in printed electronics[12]. However, lower conductivity of solution processed PEDOT: PSS, as compared to printable metal inks, limits its usability in electronic devices. Alternatively, the conductivity of PEDOT: PSS have been improved by blending it with nanoparticles[14,15]. Carbon nanotubes in a polymer matrix are known to increase the conductivity of the films due to their high aspect ratio, thereby providing highly conducting paths inside the polymer-nanotube network[16,17]. It has been previously reported that increasing the concentration of single wall carbon nanotube in poly-N-vinyl carbazole from 0.26% to 0.43% have increased the conductivity of the films by more than two orders of magnitude[18]. The electrodes patterned from thin films of carbon nanotubes and polymer can provide large active areas and high current output because of the presence of large number of carbon nanotubes for charge transport. A blend of PEDOT: PSS and carbon nanotube is an ideal candidate for replacing metal electrodes such as silver inks in printable FETs. Thin composite films of PEDOT: PSS and functionalised carbon nanotube are easily deposited using techniques like spin coating, drop casting, bar coating, slot-die printing etc. The conductivity of the electrodes based on thin films of polymer and carbon nanotube blends is also anticipated to be higher, due to the enhanced charge transport properties.

Fast and efficient electrode patterning techniques are required for industrial scalability. Laser ablation technique is an ideal candidate for patterning electrodes using PEDOT: PSS/*f*-MWCNT inks. Laser ablation is a direct structuring technique, which can completely remove both PEDOT: PSS and CNTs from the substrate, resulting in narrow electrode gaps (short channel lengths) essential for FETs. Hence laser ablation of polymer/carbon nanotube blend devices will result in a one-step patterning of electrodes for the FETs useful for upscale industrial production.

Solution processable semiconducting inorganic nanowires (NWs) are ideal candidates for active layers in printable electronic devices, offering the advantage of significantly higher charge carrier mobility as compared to printable organic semiconductors[19]. Inorganic semiconducting single crystal nanowires maintain most of their bulk single crystalline properties including very efficient charge transport and highly ordered crystal lattices. Additionally, Supercritical Fluid-Liquid-Solid (SFLS) growth method for the production of silicon and germanium nanowires is efficient, and it can offer high throughput of nanomaterials of up to few kgs/day[20]. From device fabrication point of view, SFLS grown nanowires offer compatibility with low temperature processes, due to the possibility of separating the synthesis and device fabrication procedures [21,22]. Furthermore, alignment of nanowires across the electrodes using electric field assembly technique, dielectrophoresis (DEP), results in highly ordered array of nanowires, a big advantage for transistor device[21,23].

In this work we demonstrate the feasibility of using laser patterned PEDOT: PSS/acid functionalised multiwall carbon nanotubes (f-MWCNT) composite electrodes for SFLS grown silicon nanowire field effect transistors (FETs) on flexible plastic substrates. We have shown an increase in conductivity of PEDOT: PSS after blending with f-MWCNT, and an increase in absolute workfunction values for blended samples as compared to pristine PEDOT:PSS. We demonstrate the alignment of nanowires across laser patterned PEDOT: PSS/*f*-MWCNT electrodes using dielectrophoresis technique. Finally, we illustrate fully working Si NW-FETs with laser patterned PEDOT: PSS/*f*-MWCNT electrodes on flexible substrate, with excellent gate modulation and good transistor parameters.

## Results and Discussion

### PEDOT: PSS and *f*-MWCNT composite electrodes

Functionalisation of carbon nanotubes and preparation of formulations are described in detail in Experimental sections I, II. Thin films of PEDOT: PSS and PEDOT: PSS/f-MWCNT were prepared on polyethylene terephthalate (PET) substrates, via drop casting technique, resulting in a film thickness of ~ 800nm. Such film thickness of PEDOT:PSS/f-MWCNT films makes it difficult for imaging techniques such as SEM of AFM to examine the presence of carbon nanotubes. Raman spectroscopy offers the potential to determine the presence of MWCNT in PEDOT:PSS polymer matrix by identifying characteristic Raman peaks associated with carbon nanotube vibrations. Films of PEDOT:PSS and PEDOT:PSS/f-MWCNT were examined using Renishaw system 2000 Raman spectrometer with 782nm excitation laser. Raman spectra of both samples were obtained between 95cm$^{-1}$ to 3500 cm$^{-1}$ as demonstrated in figure 1 (A). The PEDOT:PSS sample has a G-peak at 1592 cm$^{-1}$ with a D-peak at 1435 cm$^{-1}$, when compared to the MWCNT with a G-peak at 1586 cm$^{-1}$ and a significant D-peak at 1348 cm$^{-1}$. The G peak of PEDOT:PSS can be attributed to the C=C asymmetric stretching of the thiophene rings at the centre and the ends of the polymer chain, and the D-peak is due to the C-C stretching vibrations between the localised excitations[14]. The D peak for f-MWCNT is due to the defects caused by acid functionalisation, and the G-peak is due to the lattice vibration of the carbon atoms[24]. However, the PEDOT:PSS/MWCNT samples illustrated two significant D-peaks at 1359 cm$^{-1}$ and 1435 cm$^{-1}$ which correspond to the D-peak from PEDOT:PSS and MWCNT's. G-peak at 1582 cm$^{-1}$ was also observed as shown in figure 1 (A). The $I_D/I_G$ ratio was estimated to be 0.85, which gives the information about the amount of impurities in the sample. The presence of carbon nanotubes can be verified from the presence of a strong 2D peak at 2708 cm$^{-1}$ in Raman spectra of PEDOT:PSS/MWCNT sample in comparison with the PEDOT:PSS sample, where this peak is absent. 2D peak at 2708 cm$^{-1}$ is an overtone of the D peak and its existence is common in nanocarbons with sp2 hybridised carbon orbitals, such as carbon nanotubes in our sample[25]. Thus, Raman data in figure 1 (A) confirms the presence of carbon nanotubes in PEDOT:PSS / MWCNT composite film samples.

The effect of the addition of f-MWCNT into PEDOT:PSS matrix on electrical properties of films was determined by performing two-point probe current-voltage characteristics measurements of PEDOT: PSS and PEDOT: PSS/f- MWCNT composite samples. Figure 1 (B) shows the current vs voltage characteristics of PEDOT:PSS/f- MWCNT and PEDOT:PSS thin films of the same thickness. The conductivity values were extracted from figure 1 (B), PEDOT: PSS/f- MWCNT film gave higher conductivity value of $3.9 \times 10^{-3}$ Sm$^{-1}$ when compared to the conductivity value of $5 \times 10^{-7}$ Sm$^{-1}$ achieved for the pristine PEDOT:PSS film. The increase in conductivity for the PEDOT:PSS/f-MWCNT samples can be due to the more conductive PEDOT components and depletion of PSS on the surface[14], and due to the presence of highly conducting MWCNT. In addition to the current-voltage characteristic, we also determined the absolute work function of PEDOT:PSS/MWCNT and PEDOT:PSS films using Kelvin probe method. The absolute workfunction value for PEDOT:PSS was estimated to be ~ 4.9eV ± 0.1eV, whereas PEDOT:PSS/MWCNT blend films gave an absolute workfunction of ~ 5.4eV ± 0.1eV. Higher workfunction values obtained for PEDOT:PSS/MWCNT films makes them an ideal candidate for high workfunction printable inks for electrodes.

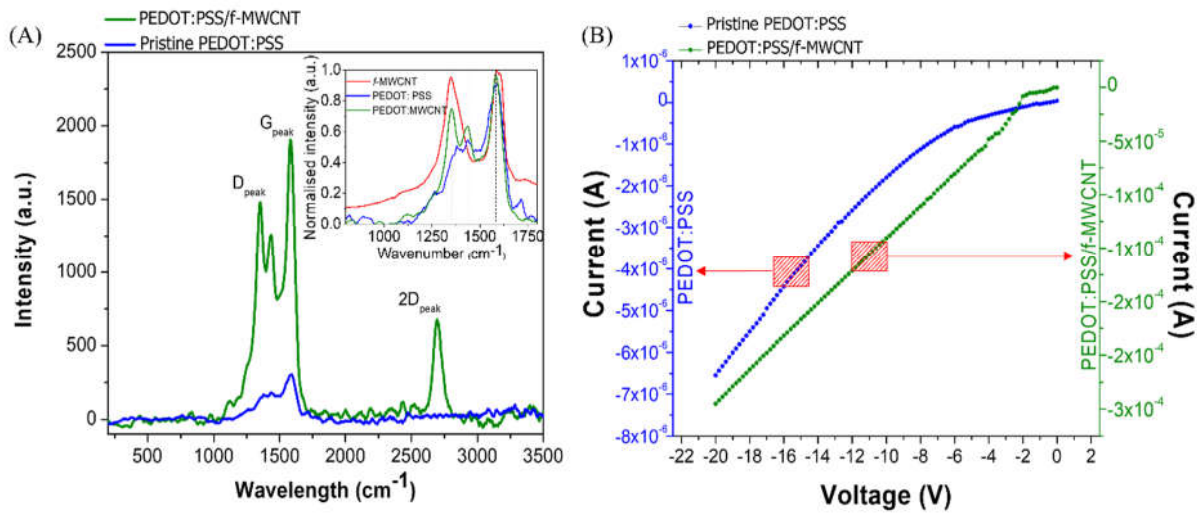

*Figure 1: Characterisation of PEDOT: PSS and PEDOT: PSS/MWCNT thin films, (A) Raman spectroscopy illustrates the presence of D and 2D peaks for PEDOT: PSS/MWCNT thin films; (B) Two point probe I-V measurements showing higher current of composite PEDOT: PSS/MWCNT thin film as compared to pristine PEDOT:PSS film.*

## Laser Patterning of PEDOT:PSS/MWCNT electrodes

Films of PEDOT: PSS/f-MWCNT (~800nm) on PET substrates were used to fabricate source/drain electrodes for transistors, via laser ablation technique. The fabrication conditions are described in Experimental section III. Figure 2 (B) shows polarized optical microscope image of a typical sample, obtained by laser ablation of PEDOT:PSS/MWCNT film, resulting in 36μm channel gap between two electrodes. To check if the conducting material was completely removed in the channel gap area, current-voltage characteristics were measured between the two electrodes patterned by laser ablation, using Keithley SCS 4200 semiconductor analyzer. A higher resistance ($\sim 10^{13}\Omega$) between the electrodes was observed after laser ablation of electrodes, as compared to un-treated PEDOT:PSS/MWCNT electrodes ($\sim 10^{5}\Omega$). High resistance in G$\Omega$ range across laser ablated electrodes illustrates the absence of PEDOT: PSS/MWCNT residue in the channel area, and an excellent isolation of the electrodes.

For FET device fabrication, solution processed silicon nanowires, with 30- 50nm diameters and few tens of microns in length as shown in figure 2 (A), were deposited to bridge the channel, using dielectrophoresis (see Experimental section IV). A dense array of silicon nanowires was observed in the channel area between the laser patterned PEDOT: PSS/MWCNT electrodes as shown in figure 2 (C).

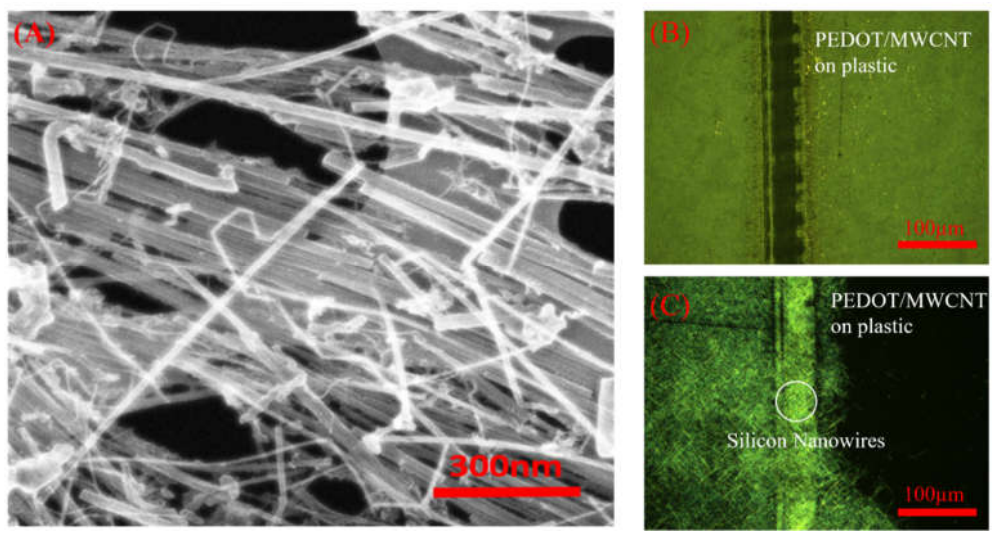

*Figure 2: (A) TEM image of SFLS grown silicon nanowires on a carbon grid, (B) Optical polarised microscope image of PEDOT: PSS/MWCNT electrodes patterned using laser ablation, (C) Dark-field polarised optical microscope image of DEP aligned silicon nanowires on laser patterned PEDOT: PSS/MWCNT electrodes.*

## Silicon nanowire FETs

Top-gate NW FETs were fabricated using the laser ablated PEDOT: PSS/f-MWCNT electrodes with DEP aligned Si-NWs (see Experimental section IV). The schematic of top-gated Si-NEW FET fabricated on top of the aligned nanowires is shown in figure 3 (A), with a channel length of ~ 36μm and channel width ~ 500μm (area of nanowire coverage).

Transfer characteristics of Si-NW FETs were obtained by sweeping the gate voltage ($V_G$) from +20V to -60V at 1V/s gate sweep rate, and measuring the drain current ($I_D$) at a constant drain bias voltage ($V_D$), as shown in figure 3 (A). From the transfer characteristics, it is observed that Si-NW FET showed p-type conduction and good modulation with negative gate voltages. The output characteristics obtained by measuring drain current for different drain voltage, at constant gate voltage as shown in figure 3 (B). An increase in drain current for different gate bias voltages in the output characteristics demonstrates good gate electrode - channel coupling. Transistor ON-state current was relatively modest, being in the range of 10s of nano- amperes, which could be attributed to a long channel length of the transistor, and also very dense coverage of nanowires. Such dense channel coverage could lead to a strong screening of the gate electric field by the nanowires, resulting in a modest charge carrier accumulation in Si nanowires channel.

The effective device mobility was calculated using the following equation.

$$\mu = \frac{L}{W \times C_{ins}} \times \frac{1}{V_{SD}} \times G_m \qquad (1)$$

Where L and W are the respective channel length and channel width, $V_{SD}$ is the drain voltage and $C_{ins}$ is the capacitance of the parylene C (~ 2.7nF/cm$^2$). Gm (~ $\partial I_D/\partial V_G$) is the transconductance of the transistor. Due to a dense 'mat' of nanowires it was not possible to calculate the exact number of nanowire crossing the channel, and thus, only effective *device* mobility was estimated.

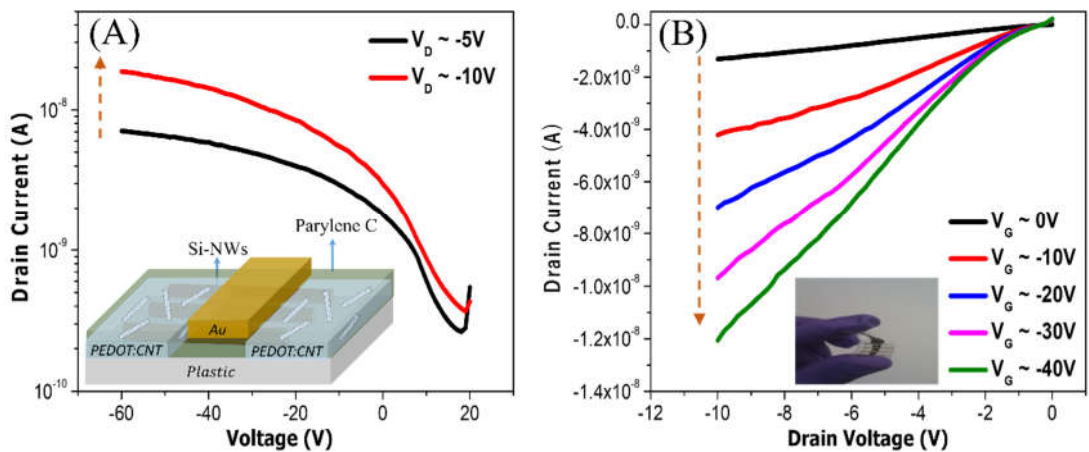

*Figure 3: Si-NW FET characteristics; (A) transfer characteristics obtained by scanning gate voltage from +20V to -60V at 1V/S, at constant drain bias voltages -5V and -10V, (B) Output characteristics obtained by scanning drain voltage from 0V to -10V, at constant gate voltages 0V, -10V, -20V, -30V and -40V. Inset shows a photographic image of FETs on flexible PET substrate.*

A transconductance value ($G_m$) of ~ 0.12nS, and an $I_{ON}/I_{OFF}$ ratio of ~ 200 were obtained from transfer characteristics, shown in figure 3(A). The effective device mobility value was calculated to be ~ 6.4×10$^{-4}$cm$^2$/Vs using equation 1.

## Conclusions

In summary, we have developed a high workfunction, printable PEDOT:PSS/MWCNT composite ink for source-drain electrodes fabrication in nanowire FETs. Importantly, this seems to be the only low-cost inks with high workfunction compatible with silicon nanoparticles valence band edge (5.1eV) for p-type transport. Gold nanoparticle inks, although, are significantly more expensive, and their sintering temperature usually exceeds glass transition temperature of flexible PET substrates, making them incompatible with low temperature printed electronics.

We have illustrated a higher conductivity value of ~ $3.9 \times 10^{-3}$ Sm$^{-1}$ for PEDOT: PSS/MWCNT films as compared to the pristine PEDOT: PSS film (~ $5 \times 10^{-7}$ Sm$^{-1}$). Furthermore, it was also observed that the work function value of PEDOT: PSS films has increased from ~ 4.9eV ± 0.1eV to ~ 5.4eV ± 0.1eV with the addition of MWCNTs. Increase in workfunction values can help to improve the charge injection from the electrodes into the semiconductor for the p-type FETs that are based not only silicon nanowires, but also high ionization potential semiconducting small molecules and polymers.

The source/drain electrodes were patterned from PEDOT: PSS/MWCNT films using laser ablation process on plastic substrates. We have routinely obtained ~36μm gap between the S/D electrodes. Top gate Si-NW FETs were fabricated from laser patterned composite electrodes. Devices showed an $I_{ON}/I_{OFF}$ ratio > $10^2$, with an effective device mobility values of ~ $6.4 \times 10^{-4}$ cm$^2$/Vs. Overall, silicon nanowire FETs with laser ablated PEDOTPSS/MWCNT electrodes demonstrated the feasibility of using high workfunction composite PEDOT:PSS/MWCNT inks for nanowire transistors. In is envisaged that laser ablation of transistor electrodes can reduce the overall cost of printed electronic devices.

# Experimental

### I. Functionalisation of carbon nanotubes

Functionalisation of the carbon nanotubes is an important process to: (i) make CNTs soluble in common solvents used for conducting polymers, and (ii) prevent the carbon nanotubes from agglomeration and forming bundles in the thin films, and also maintaining good dispersion within the polymer matrix. MWCNTs were purchased from Sigma Aldrich with the dimensions 7-15nm × 3-6nm × 0.5-200 $\mu$m, and were treated using acid functionalisation technique. 500mg of MWCNT's were mixed with 30ml of concentrated sulphuric acid and concentrated nitric acid mixture (3:1). The solution was sonicated in an ultrasonic bath for 10 minutes, and refluxed over an oil bath for 1 hour at 130ºC. The mixture was allowed to cool down and was diluted to 80ml using MilliQ deionised water, and was then transferred to two 50ml centrifuge tubes. The diluted solution was centrifuged at 8700rpm for 25min. A brown colored supernatant was observed which was then removed using a homemade vacuum pump, resulting in a black precipitate. The black precipitate was then diluted using MilliQ deionised water and the precipitate was suspended using a vortex mixer. The process was repeated twice to remove all the concentrated acid used for functionalisation, resulting in a black suspension. The solution was filtered using 0.1 $\mu$m polycarbonate membrane filter, washed with deionised water until the required pH (~7) level was reached, which was then finally washed with absolute ethanol (Fisher AR grade). The functionalized COOH-MWCNTs were dried in vacuum desiccator and stored as dry powder. For making formulations, dry COOH-MWCNTs were added into 50% ethanol/deionised water to a concentration of 0.5mg/ml.

## II. Preparation of the formulation: f-MWCNT / PEDOT

The solution of 0.5mg/ml COOH-MWCNT was filtered using a 0.1 $\mu$m polycarbonate membrane filter. 30ml of 0.5mg/ml was taken from the solution. 1.3% poly (3, 4-ethylenedioxythiophene) poly (styrenesulphonate) (PEDOT:PSS) purchased from Clevios was filtered using a 0.22 $\mu$m filter followed by sonicating in an ultrasonic bath for 4 minutes. 1ml of PEDOT: PSS was mixed with 1ml of COOH-MWCNTs and transferred into a vial. The process was repeated so as to get 3 vials of PEDOT: PSS/MWCNTs with 1:1 ratio.

## III. Laser patterning of electrodes

The Excimer Compex 205 (Lamda Physik) laser operating at 248 nm wavelength with repetition rate of 80Hz and a spot beam size (after focusing lens) of 36μm was used to pattern PEDOT:PSS/MWCNT electrodes. An 800nm thick film of PEDOT:PSS/MWCNT was prepared on the PET substrate using drop casting technique, and oven dried at 100˚C for 15min. The laser beam was homogenized to provide uniform intensity distribution in the beam cross-section. A fixed pattern to be drawn on the sample using the laser beam was pre-programmed. The sample was placed on an X-Y translation stage which was programmed to move at a speed of 10mm/sec. The beam was focused on the thin film and the programmed pattern was drawn on the film. Laser energy of 44mJ with a power density of 339.5 mJ/cm$^2$ was required to remove all the material from the channel area for a thickness of 800nm. The channel length obtained was 36μm measured using the optical polarization microscope, and the channel depth was found out to be 855nm using profilometer. The profilometer measurement suggests that the laser beam is penetrating almost 55nm into the substrate, after removing the PEDOT: PSS/MWCNT thin film.

## IV. Fabrication of top-gated silicon nanowire transistor

SFLS synthesised silicon nanowires with diameters of 30nm to 50nm and a few tens of microns in length were used to form semiconducting layer between laser patterned electrodes. Silicon nanowires were dispersed in anisole, and then aligned using dielectrophoresis method by applying AC voltage of 10Vpp to the PEDOT/MWCNT electrodes and then drop-casting nanowire dispersion on top. Excess anisole was removed, samples were dried, and 1μm thick layer of parylene C was deposited on top of the electrodes as the dielectric layer[23], and 50nm thick gold gate electrode was evaporated using a Kurt Lesker thermal deposition system on top of the dielectric layer through a shadow mask. The transistor characteristics were measured in ambient air, at room temperature, using Agilent 4155C semiconductor parameter analyzer.

# Acknowledgements

The authors would like to thank Dr. Simon Henley, (ATI, University of Surrey, UK) for his valuable insight and advice on laser ablation technique. The author would also like to extend their gratitude to Prof. Brian A. Korgel (University of Texas, Austin, USA) for providing the SFLS grown silicon nanowires.